\documentclass[aps,prc,twocolumn,superscriptaddress,showpacs,showkeys]{revtex4-1}

\usepackage{graphicx}
\usepackage[colorlinks=true,linkcolor=blue,bookmarks=true,urlcolor=blue,citecolor=blue,dvipdfmx]{hyperref}

\begin{document}


\title{
Deformation and cluster structures in $^{12}$C studied with configuration mixing 
using Skyrme interactions
}


\author{Y. Fukuoka}
\email[]{fukuoka@nucl.ph.tsukuba.ac.jp}
\affiliation{Graduate School of Pure and Applied Sciences, University of Tsukuba, Tsukuba 305-8571, Japan}

\author{S. Shinohara}
\affiliation{Kudoh Patent Office, 6-24-10 Minami-Oi, Shinagawa, Tokyo 140-0013, Japan}

\author{Y. Funaki}
\affiliation{RIKEN Nishina Center, 2-1 Hirosawa, Wako 351-0198, Japan}

\author{T. Nakatsukasa}
\affiliation{RIKEN Nishina Center, 2-1 Hirosawa, Wako 351-0198, Japan}
\affiliation{Center for Computational Sciences, University of Tsukuba, Tsukuba 305-8577, Japan}

\author{K. Yabana}
\affiliation{Graduate School of Pure and Applied Sciences, University of Tsukuba, Tsukuba 305-8571, Japan}
\affiliation{Center for Computational Sciences, University of Tsukuba, Tsukuba 305-8577, Japan}

\date{\today}

\begin{abstract}
We report an investigation of the structure of $^{12}$C nucleus 
employing a newly developed configuration-mixing method.
In the three-dimensional coordinate-space representation,
we generate a number of Slater determinants with
various correlated structures using the imaginary-time algorithm. 
We then diagonalize a many-body Hamiltonian with the Skyrme 
interaction in the space spanned by the Slater determinants 
with parity and angular momentum projections.
Our calculation reasonably describes the ground and excited states
of $^{12}$C nucleus, both for shell-model-like and cluster-like 
states. The excitation energies and transition strengths of 
the ground-state rotational band are well reproduced.
Negative parity excited states, $1_1^-$, $2_1^-$, and $3_1^-$, are
also reasonably described. The second and third $0^+$ states,
$0_2^+$ and $0_3^+$, appear at around 8.8 MeV and 15 MeV, 
respectively. The $0_2^+$ state shows a structure consistent 
with former results of the $\alpha$-cluster models,
however, the calculated radius of the $0_2^+$ state is smaller
than those calculations. The three-$\alpha$ 
linear-chain configuration dominates in the $0_3^+$ state. 
\end{abstract}

\pacs{21.10.Dr,27.20.+n}

\maketitle

\section{INTRODUCTION}

Light nuclei show a variety of structures in the ground and
excited states characterized by different correlations.
The nuclear shell model usually provides a reasonable description
for the ground and low-lying states. However, it is difficult for
the shell model to describe well-developed cluster states
which appear in excited states of light nuclei.
The appearance of the cluster structures 
is intimately related to the small separation energy of 
clusters. This is called the threshold rule and has been 
described by the so-called Ikeda diagram \cite{KIkeda1968}.

For theoretical descriptions of cluster states, microscopic and 
semi-microscopic cluster models have been extensively developed
in the past \cite{PTPS.62.1,PTPS.E68.464,KIkeda1980,YFujiwara1980}. 
The resonating group method (RGM) 
\cite{JAWheeler1937_1,JAWheeler1937_2} 
assumes a product form for the wave function composed of the
internal wave functions of clusters and the inter-cluster wave 
function, taking fully account of the antisymmetrization.
The generator coordinate method (GCM) was also successfully applied to
various cluster motions assuming harmonic oscillator 
shell-model wave functions for clusters.
as well \cite{DMBrink1968}. In these microscopic models, the existence 
of clusters is assumed from the beginning. 
To understand mechanisms of emergence and disappearance
of cluster structures, one should start with a model which does 
not assume any existence of clusters. Studies with the antisymmetrized 
molecular dynamics (AMD) method \cite{YKEnyo1995_1,YKEnyo1995_2,Kanad2007}
have contributed to substantial advances in this direction. 
In most calculations with microscopic cluster models and AMD,
effective nucleon-nucleon forces are used. In the Fermionic Molecular
Dynamics (FMD) method which is closely related to the AMD, a more realistic
force produced by the unitary correlation operator method has been 
employed \cite{TNeff2004}.

In last two decades, there have been significant advances in
theoretical descriptions of light nuclei
starting with realistic nucleon-nucleon force, so-called the
ab-initio approach.
The Green Function Monte Carlo (GFMC) approach has been 
successful to describe the ground and the low-lying excited states 
of light nuclei \cite{JCarlson1987}. The GFMC
calculation describes the two $\alpha$ cluster structure of $^8$Be
in the ground state \cite{RBWiringa2000}. The no-core shell model (NCSM) 
has also been successful for the description of ground and 
low-lying excited states \cite{PNavratil2000,PNavratil2009}. 
However, descriptions of cluster states in the NCSM have not yet
been satisfactory.
Recently there are a number of attempts for the ab-initio
description of cluster structures in excited states. For example,
a lattice calculation for the Hoyle state has been reported
\cite{PhysRevLett.106.192501}. The no-core shell model combined
with Monte-Carlo basis generation method has also been applied
\cite{PhysRevC.86.054301}.

There are two important ingredients in the ab-initio descriptions of
nuclear structures. One is to start with a
Hamiltonian with a realistic nucleon-nucleon force that has a
short-range repulsive core.
The other is to obtain fully convergent solutions for
the many-body Hamiltonian. Since cluster structures
are characterized by long-range spatial correlations,
simultaneous descriptions of both long- and short-range
correlations are required in the ab-initio calculations
of cluster states. This makes the problem computationally very challenging.

In this paper, we focus on the latter aspect of
the above-mentioned problem, namely, on obtaining fully convergent
solutions for a given many-body Hamiltonian, taking into account
a variety of long-range correlations. We start
not with a realistic nucleon-nucleon force but with
an empirical effective interaction, the Skyrme force.
We use a newly developed method which was
reported previously \cite{SShinohara2006}. We apply the method 
to $^{12}$C nucleus, then, examine descriptions of cluster structures.
Since the Skyrme interaction is determined so as to
reproduce nuclear properties of wide mass region,
our calculation contains no empirical parameters specific
to $^{12}$C.

Among light nuclei, the $^{12}$C nucleus is one of the most 
interesting systems for the reasons described below. 
In the $jj$ coupling shell-model picture, the ground 
state wave function should be dominated by the $p_{3/2}$ closed 
shell configuration. Indeed, the self-consistent Hartree-Fock 
solutions with most Skyrme interactions show a spherical shape 
with the $p_{3/2}$ closed shell configuration. 
However, the $^{12}$C nucleus is known to show a rotational 
band structure built on the ground state, indicating a deformed 
intrinsic shape in the ground state. In excited states, 
a variety of cluster structures are known to appear. 
The $0_2^+$ state just above the three-$\alpha$ decay threshold 
is an important resonant state for the triple-$\alpha$ fusion 
reaction, and is known as the Hoyle state \cite{Hoyle1954}. 
It has been found recently that this state is well described by 
a Bose condensed wave function of three-$\alpha$ particles 
which is called the THSR wave function
\cite{ATohsaki2001,YFunaki2003}. 
The appearance of three-$\alpha$ linear-chain structure in excited
states was suggested by Morinaga in 1966 \cite{Morinaga1966}. 
Recent microscopic cluster models predict that the $0_3^+$ 
state is a candidate for the linear-chain like structure 
\cite{Kanad2007,TNeff2004}.

This paper is organized as follows. In section~\ref{s:formulation}, we explain 
our method. In section~\ref{s:results}, we first show that we obtain convergent
results for some low-lying states for the Skyrme Hamiltonian. 
We then describe in detail the calculated results for $^{12}$C. 
In section~\ref{s:mix_conf}, we compare our results with conventional microscopic
cluster model calculations by introducing cluster wave functions 
in our calculation. In section~\ref{s:summary}, a summary is presented.

\section{Formulation}
\label{s:formulation}

In this section, we present our formalism which was proposed
in Ref. \cite{SShinohara2006}.
It is composed of three steps: 
We first generate a number of Slater determinants (SDs) in a stochastic way. 
These SDs are expected to span a sufficiently large model space 
to describe excited states with various cluster structures as well as
low-lying states with shell-model-like structures.
We then perform parity and angular momentum projections for the SDs. 
Finally we superpose them to diagonalize the Skyrme Hamiltonian. 
Below we describe these three steps of our formalism in order.

Before presenting our formalism, we briefly describe
numerical aspects in the present method.
For the energy functional, we employ the SLy4 parameter set of the
Skyrme interaction unless otherwise specified.
To describe single-particle orbitals of the SDs, 
we employ a representation of the three-dimensional (3D) Cartesian grid. 
This representation allows us a flexible 
description of single-particle orbitals in arbitrary nuclear shapes.
The grid spacing is taken to be 
$\Delta x=\Delta y=\Delta z=0.8$ fm,
and all the grid points inside a sphere of radius, $R_{\rm max}=8.0$ fm, 
are adopted.

\subsection{Generation of Slater determinants}\label{s:Generation_SD}

\begin{figure*}[Htb]
  \begin{center}
    \includegraphics[]{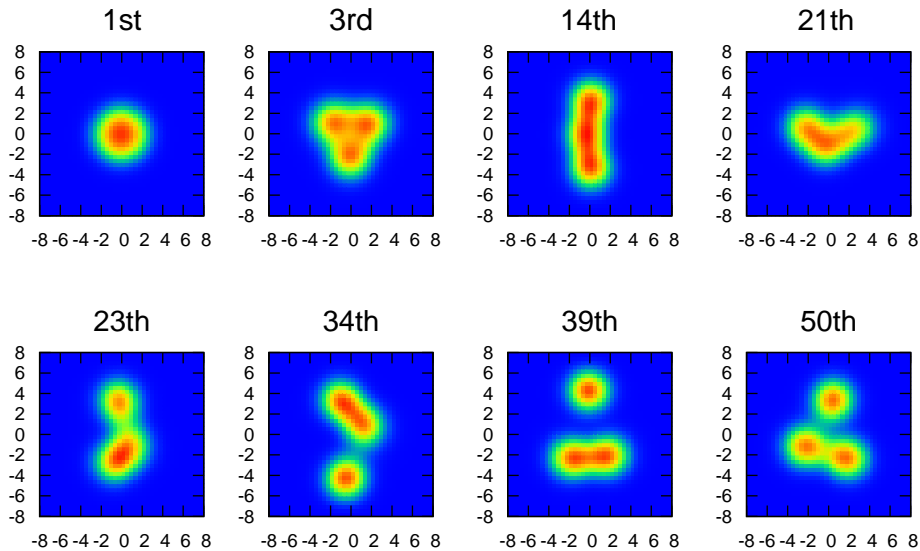}
  \end{center}
  \caption{Contour plots of nuclear densities of the stored SDs for $^{12}$C. 
    A sequential number of the SD is indicated in the top of the panel.
    Units of vertical and horizontal axes are fm.}
  \label{wf_imag}
\end{figure*}

The first step is to generate and select a sufficient number of SDs,
$\{ \Phi_i \}$ ($i=1,\cdots, M$),
which span a large model space to describe various kinds of
long-range correlations.
To this end, we make use of the imaginary-time 
method with the Skyrme interaction. The imaginary-time method is 
usually used to obtain a ground-state solution in the mean-field 
calculation. Here, we use it for generation of many kinds of
collective surfaces.

We start with initial SDs whose single-particle orbitals
are described by the Gaussian wave packets,
\begin{equation}
  \phi_i(\vec{r},\sigma )=e^{-|\vec{r}-\vec{R_i}(\sigma)|^2/a^2}. 
  \label{initSD}
\end{equation}
The centers of the Gaussian wave packets, $\vec R_i(\sigma)$, are set by 
random numbers generated under the condition,
\begin{equation}
  |\vec R_i(\sigma)|<R_{\rm max}-1 \hspace{2mm} {\rm fm}.
\end{equation}
The width parameter $a$ is taken to be 2 fm.

Then, 
we start the imaginary-time iterations with these initial SDs.
During the imaginary-time iterations, we set constraints
to place the center of mass at the origin and to make the
principal axes of nuclei parallel to the Cartesian axes.
After the sufficient number of iterations, it reaches the 
self-consistent ground-state solution, namely Hartree-Fock (HF) state.
However, before reaching the HF solution,
there appear a number of configurations which show various cluster
structures and other configurations important for low-energy nuclear
dynamics. Thus, we pick up and store these SDs which will be used for the 
configuration mixing calculation. We repeatedly perform the 
imaginary-time iterations starting with different initial SDs 
to obtain a sufficient number of SDs (typically 50) of many kinds of
correlations.

In order to span a wide model space by the limited number of SDs,
the actual selection of the SDs is achieved as follows:
The first SD adopted in the basis set $\{ \Phi_i \}$ is
the HF state, $\Phi_1=\Phi_{\rm HF}$.
The second and following SDs are generated as follows:
During the imaginary-time iteration, 
the energy expectation value 
decreases monotonically. We examine the energy expectation value 
in a regular interval, typically every 100 iterations. We do not adopt 
the SD until the energy expectation value is less than 30 MeV 
above the HF ground-state energy. After the energy expectation value 
becomes less than this threshold, we examine 
the overlap between a current SD and previously selected ones.
We denote the SDs which have been already
selected as $\Phi_i(i=1,\cdots, N)$, and denote the current SD 
as $\Phi_{c}$. We calculate the quantities
\begin{equation}
  \frac{\big|\big<\Phi_i|\hat{P}^\pi \hat R_n|\Phi_{c}\big>\big|}
  {\sqrt{\big|\big<\Phi_i|\hat P^\pi|\Phi_i\big>\big|}\sqrt{\big|\big<\Phi_{c}|\hat P^\pi|\Phi_{c}\big>\big|}}
  \hspace{2mm} (i=1,\cdots,N),
\end{equation}
where $\hat P^\pi$ is the parity projection operator and 
$\hat R_n (n=1,\cdots,24)$ are operators of rotations and inversions
which may be easily achieved by the changes of the coordinate axes. 
If the maximum value of the above overlap values between $\Phi_{c}$
and $\Phi_i(i=1,\cdots,N)$ is less than 0.7, we adopt the SD 
$\Phi_{c}$ as a new member, $\Phi_{N+1}=\Phi_{c}$.
When the imaginary-time 
iteration reaches the self-consistent solution, we generate a new 
initial SD of the form of Eq.~(\ref{initSD}) and again start
the imaginary-time iterations. 
We then repeat 
the procedure to select a new SD. We repeat the procedure until a 
sufficient number of SDs is obtained. Typically a few SDs are 
adopted during one imaginary-time iterations. As the number of stored SDs 
increases, it is more and more difficult to find the new SD which 
satisfies the overlap condition.
In this way, we store SDs 
which include important correlation effects and which are linearly 
independent to each other as much as possible.

We show density distributions of several SDs generated in this
procedure in Fig.~\ref{wf_imag}. The numbers assigned 
to the figures, 1, 3, 14, 21, 23, 34, 39, and 50, simply indicates the
adopted ordering.
The first one which shows a spherical shape 
is the HF solution for the ground state. Other SDs in Fig.~\ref{wf_imag}
show a variety of cluster structures. For example, 
$\Phi_3$ shows an equilateral triangular three-$\alpha$ structure, 
$\Phi_{14}$ shows a three-$\alpha$ linear-chain, and $\Phi_{23}$ shows 
a $^8$Be+$\alpha$ like structure. We thus observe that the present 
procedure efficiently produces SDs with various cluster structures
in an automatic manner.

\subsection{Projections of parity and angular momentum}

The SDs prepared by the method in Sec. \ref{s:Generation_SD} are,
in general, not eigenstates of parity and angular momentum. 
To calculate matrix elements between eigenstates of parity and
angular momentum, we apply the projection method. The projection 
operators are given as usual by
\begin{eqnarray}
  \hat P^\pi&=& \frac{1}{2}(1+(-1)^\pi \hat P_r), \\
  \hat P^J_{MK}&=&\frac{2J+1}{8\pi^2}\int d\Omega D^{J*}_{MK}(\alpha, \beta, \gamma)
  \hat R(\alpha, \beta, \gamma),
\end{eqnarray}
where $\hat P_r$ is the space inversion operator, 
$\hat R(\alpha,\beta,\gamma)$ is the rotation operator for the Euler 
angles, $\alpha$, $\beta$, and $\gamma$, and $D^{J}_{MK}$ is the 
Wigner's $D$-function defined by
\begin{eqnarray}
  \hat R(\alpha, \beta, \gamma) &=& e^{-i\alpha \hat J_z}e^{-i\beta \hat J_y}e^{-i\gamma J_z},\\
  D^J_{MK}(\alpha, \beta, \gamma) &=& e^{-i\alpha M}d^J_{MK}(\beta)e^{-i\gamma K},
\end{eqnarray}
where $J$, $M$, and $K$ are the total angular momentum, its projection 
onto the laboratory $z$-axis, and its projection onto the body-fixed $z$-axis, 
respectively.

We define the norm and Hamiltonian matrix elements between the projected 
SDs $\big|\Phi_i\big>$ and $\big|\Phi_j\big>$ as
\begin{eqnarray}
  n^{J\pi}_{iK,jK'} \equiv&& \frac{2J+1}{8\pi^2}\nonumber\\
   \times \int \hspace{-1.5mm}d\Omega&& D^{J*}_{KK'}(\Omega) 
   \big<\Phi_i\big|e^{-i\alpha \hat J_z}\hat P^{\pi}e^{-i\beta \hat J_y}
   e^{-i\gamma \hat J_z}\big| \Phi_j\big>,
\label{norm}
\end{eqnarray}
\begin{eqnarray}
  h^{J\pi}_{iK,jK'} \equiv&& \frac{2J+1}{8\pi^2}\nonumber\\
   \times \int \hspace{-1mm}d\Omega&& D^{J*}_{KK'}(\Omega)
   \big<\Phi_i\big|e^{-i\alpha \hat J_z}\hat H \hat P^{\pi}
   e^{-i\beta \hat J_y}e^{-i\gamma \hat J_z}\big| \Phi_j\big>.
\label{hamil}
\end{eqnarray}
Here, we use the formula,
\begin{equation}
\label{PP}
\hat{P}^{J \dagger}_{MK} \hat{P}^{J}_{MK'} = \hat{P}^{J}_{KK'}.
\end{equation}
In Eqs. (\ref{norm}) and (\ref{hamil}),
we need the rotation of wave functions.
It is achieved by successive operations of small-angle
rotations. For example, the rotation of a wave function $\phi$ over 
an angle $\gamma$ around the $z$-axis is achieved by successive 
rotations of a small angle, $\Delta \gamma = \gamma / N_{\gamma}$,
$N_{\gamma}$ times,
\begin{equation}
  e^{-i\gamma\hat j_z}\big|\phi\big>
  =\big( e^{-i\Delta\gamma\hat j_z}\big)^{N_{\gamma}}\big|\phi\big>,
\end{equation}
To achieve the small angle rotation, we employ the Taylor expansion method.
\begin{equation}
  e^{-i\Delta\gamma\hat j_z} \big|\phi\big>
\approx \sum^{N_{\rm max}}_{k=1}\frac{(-i\Delta\gamma\hat j_z)^k}{k!} \big|\phi \big>.
\end{equation}
Typically, we take $N_{\rm max} = 4$ and $\Delta\gamma=\frac{\pi}{90}$.

The integrals over Euler angles in Eqs.~(\ref{norm}) 
and (\ref{hamil}) are evaluated as follows:
Those over $\alpha$ and $\gamma$ are achieved by the trapezoidal 
rule, taking 18 uniform grid points for $[0,2\pi]$. The integral 
over $\beta$ is achieved with the Gauss-Legendre quadrature, 
taking 30 grid points for $[0,\pi]$.

\subsection{Configuration mixing}\label{s:CM}

The final procedure is diagonalization of the many-body Hamiltonian in 
the space spanned by the selected SDs. 
In Sec. \ref{s:Generation_SD}, the SDs have been screened by their linear
independence. However,
calculating eigenvalues of the norm matrix for the SDs after the
parity and angular
momentum projections, we find a number of eigenvalues very 
close to zero or even slightly negative. 
The norm matrix is positive definite by definition.
However, since we make numerical approximations in evaluating the norm
matrix, it could contain negative eigenvalues.
The approximations include use of the formula for the product of
the projection operators, Eq. (\ref{PP}),
which is no longer exact if the integral over Euler angles
is evaluated by the numerical quadrature.
We also employ the 3D Cartesian grid representation for the orbitals
in which the rotational symmetry holds only approximately.

Inclusion of those 
configurations of very small norm eigenvalues would lead to
numerical difficulties.
Therefore, we reduce the number of configurations
according to the following procedures.
First, we perform diagonalization in the $(2J+1)$-multiplet with
different $K$ quantum numbers.
\begin{equation}
  \sum_{K'} n^{J\pi}_{iK,iK'}v^{J\pi, i\nu}_{K'}
  =e^{J\pi}_{i\nu} v^{J\pi, i\nu}_K,\hspace{3mm} (\nu=1,\cdots,2J+1)
\end{equation}
where $e^{J\pi}_{i\nu}$ and $v^{J\pi, i\nu}_K$ are eigenvalues and 
eigenvectors of the norm matrix, $n^{J\pi}_{iK,iK'}$. 
We then construct a space spanned by the eigenvectors with the 
eigenvalue $e^{J\pi}_{i\nu} > 10^{-2}$, and
define the normalized basis functions 
\begin{equation}
  \big|\Phi^{J\pi}_{i\nu}\big>\equiv\frac{1}{\sqrt{e^{J\pi}_{i\nu}}}
  \sum_K v^{J\pi, i\nu}_K\hat P^J_{MK}\hat P^{\pi}\big|\Phi_i\big>.
  \label{eq:trans}
\end{equation}
After achieving the above procedure for all SDs,
we define the following norm matrix between basis functions
belonging to the different SDs,
\begin{eqnarray}
  \tilde n^{J\pi}_{i\nu,j\nu'}\equiv 
  \big< \Phi^{J\pi}_{i\nu}\big| \Phi^{J\pi}_{j\nu'} \big>
\end{eqnarray}
Using this matrix, we examine the linear independence of the basis functions
and reduce the number of basis as follows.
\begin{enumerate}

\item  Calculate eigenvalues of $2 \times 2$ matrices composed of every
possible pair of basis functions, $(i_1\nu_1)$ and $(i_2\nu_2)$.
If the smaller eigenvalue is less than $10^{-3}$, 
we remove the basis function with a smaller $e^{J\pi}_{i\nu}$.

\item Calculate the eigenvalues of $3 \times 3$ matrices composed of 
three basis functions $(i_1\nu_1)$, $(i_2\nu_2)$ and $(i_3\nu_3)$. 
If the smallest eigenvalue is less than $10^{-3}$, we remove one of 
the three basis functions in the following procedure. We calculate
eigenvalues of three $2\times 2$ submatrices composed of all possible
pairs of these three states,
to find the pair whose smaller
eigenvalue is the largest among the three.
Then we remove one of the basis functions of $(i_1\nu_1)$, $(i_2\nu_2)$ 
and $(i_3\nu_3)$ which does not belong to that pair. 
We repeat the procedure for all possible combinations of three basis functions.

\item Finally we calculate eigenvalues of the norm matrix 
$\tilde n^{J\pi}_{i\nu, j\nu'}$ with basis functions which survived in
the previous two screening steps.
If we find the eigenvalue smaller than $10^{-3}$, we remove one basis
function in the following way.
Denoting the number of basis functions as $N$, we construct
the $(N-1)\times (N-1)$ submatrices
removing one basis function $\Phi_{i\nu}^{J\pi}$ from the $N$ basis.
Apparently, $N$ different choices of $(i\nu)$ are possible.
We then calculate the smallest eigenvalue of the $(N-1)\times(N-1)$
submatrix, $\lambda_{\rm min}^{(i\nu)}$.
Among $\lambda_{\rm min}^{(i\nu)}$ with different $(i\nu)$,
we find the largest one, $\lambda_{\rm min}^{(i'\nu')}$, and
remove the basis function $\Phi_{i'\nu'}^{J\pi}$.
In this way, the number of basis $\{ \Phi_{i\nu}^{J\pi} \}$
is reduce by one, from $N$ to $N-1$.
 We repeat this process until the smallest eigenvalue 
of the norm matrix becomes larger than $10^{-3}$. 

\end{enumerate}

After removing the overcomplete basis functions in this procedure, 
we achieve the configuration mixing calculation.
Denoting the $n$th energy eigenstate as
\begin{eqnarray}
  \big| \Psi^{J\pi}_n\big>&=&\sum_{i\nu} f_{i\nu}^{J\pi, n} 
\big|\Phi^{J\pi}_{i\nu}\big> \label{e:total_wf},
\end{eqnarray}
the generalized eigenvalue equation for the energy eigenvalues
$E_n^{J\pi}$ and the coefficients $f_{i\nu}^{J\pi,n}$ 
is given by
\begin{eqnarray}
  \sum_{j\mu}\big\{ \tilde h_{i\nu j\mu}^{J\pi}-E_{n}^{J\pi}\tilde n_{i\nu j\mu}^{J\pi}\big\} f_{j\mu}^{J\pi,n} & = & 0, \label{eq:f_jmul}
\end{eqnarray}
where $\tilde h^{J\pi}_{i\nu,j\nu'}$ is defined by
\begin{eqnarray}
  \tilde h^{J\pi}_{i\nu,j\nu'}\equiv 
  \big< \Phi^{J\pi}_{i\nu}\big| \hat H \big| \Phi^{J\pi}_{j\nu'} \big>.
\end{eqnarray}

\section{RESULTS}\label{s:results}

\subsection{Convergence of results: Statistical treatment}\label{s:conv_el}

In principle, the configuration mixing calculation with 
a sufficient number of SDs should provide unique and convergent 
energy levels. However, as we described in Sec.~\ref{s:formulation},
superposing a large number of non-orthogonal 
SDs causes numerical difficulties.
In the present calculations, we adopt 50 SDs for the 
configuration mixing calculation. It is difficult to increase this number.
Further increase of the number of SDs may produce
unphysical solutions
whose energies are a few tens of MeV lower than the ground state 
energy of the HF solution.
This is possibly due to accumulations of numerical errors by
the violation of rotational symmetries in the 3D grid representation,
insufficient accuracy in numerical quadratures, and so on.

Because of the difficulty, we will not attempt to 
examine the convergence of the energy levels by increasing the 
number of SDs.
Instead, we prepare several sets of the SDs and
calculate energy levels for each set.
If the calculated energy levels are close to each other 
among the different sets of SDs, one may conclude that the calculated 
energy levels are reliable. 
In practice, we prepare ten sets, each of which is composed of
50 SDs. The ten sets of SDs are prepared in the procedure 
explained in Sec.~\ref{s:Generation_SD}. 
Different seeds for the random numbers, which are
used to prepare initial states in 
Eq.~(\ref{initSD}), are employed to generate the different sets.

\begin{figure*}[Hbt]
  \includegraphics{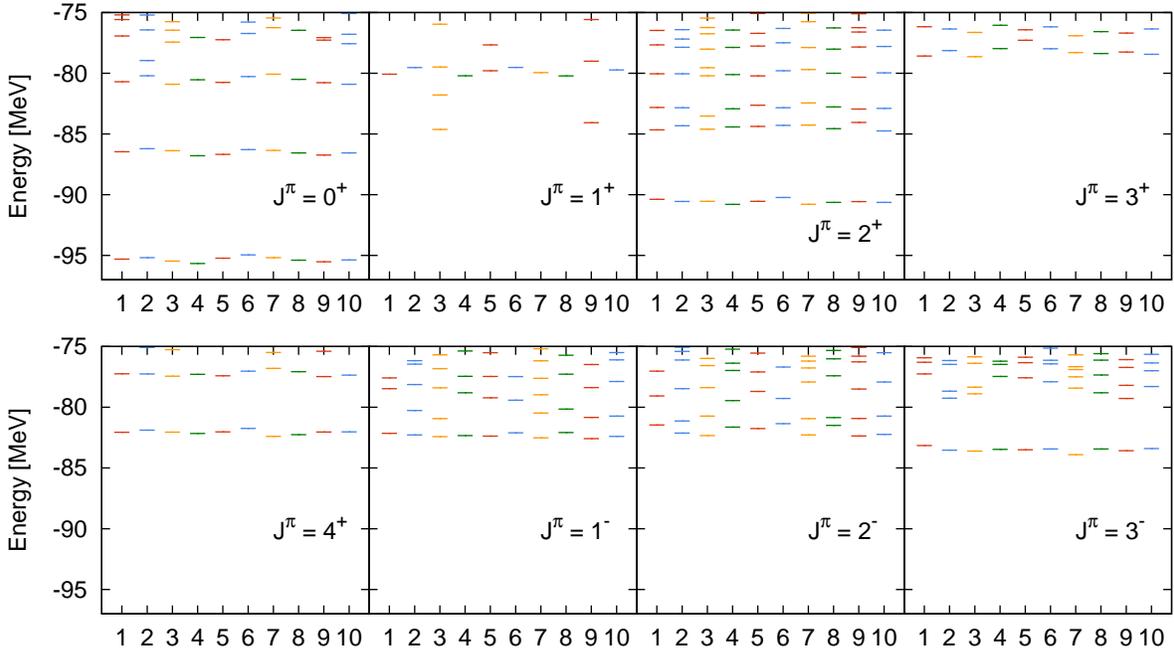}
  \caption{Energy levels of $^{12}$C nucleus for 
    $J^{\pi}=0^+, 1^+, 2^+, 3^+, 4^+, 1^-, 2^-$ and $3^-$.
    Calculations employing ten different sets of SDs are
    shown. See the text for details.}
  \label{f:ene_dif}
\end{figure*}

In Fig.~\ref{f:ene_dif}, we show the energy levels of $^{12}$C 
nucleus for the ten sets calculated in the procedure explained in 
Sec.~\ref{s:formulation}. The energy levels are shown for 
$J^{\pi}=0^+, 1^+, 2^+, 3^+, 4^+, 1^-, 2^-$ and $3^-$. 

Let us first examine calculated energy levels of $J^{\pi}=0^+$.
The lowest
level is located around -95 MeV. The difference of energies
among the ten sets is smaller than 1 MeV. 
The second excited state appears around -86 MeV.
The difference among the ten sets is again around
1 MeV. The third excited state appears around -81 MeV.
We may state that the energies of these three lowest
states are calculated reliably, since the variation
is rather small. However, energies of
fourth excited state do not show a good convergence.
For example, the energy levels of 2nd set
give the energy at around -79 MeV, close to the 3rd state.
However, the energy in the 7th set is substantially high,
approximately -76 MeV. We thus conclude that we can obtain reliable 
excitation energies and wave functions for the lowest three levels
for $J^\pi=0^+$.

The energy levels of $J^{\pi}=2^+$ in Fig.~\ref{f:ene_dif} indicate
that the energies of the lowest four states 
are reliable with a small variation.
For $J^{\pi}=3^+$ and $4^+$ states, the lowest two states 
may be reliable.
However,
the calculated energies of $J^{\pi}=1^+$ levels show
strong variation among the ten sets even for the lowest level. 
This may be due to the fact that the $J^{\pi}=1^+$ components of
the wave function disappear
in early stages of the imaginary-time iterations,
since components of high-lying levels 
quickly decay by the imaginary-time propagation.
For the negative-parity levels, only the lowest level for each
$J^{\pi}$ may be reliable. The energies of second lowest 
levels show a large variation among the ten sets for
$J^\pi=1^-, 2^-$ and $3^-$. 

For physical quantities such as energies, transition strengths,
and radii, we calculate statistical averages and standard 
deviations among the ten sets.
The average energy for the $n$-th level of $J^{\pi}$ state
is defined by
\begin{equation}
  \overline E^{J\pi}_n= \frac{1}{N_s}\sum_{i=1}^{N_s} E^{J\pi}_{n,i}
  \label{e:ave}
\end{equation}
where $i$ specifies a set among the ten sets, $N_s=10$. 
The average excitation energies are calculated as
$\overline E^{J\pi}_n - \overline E^{J\pi}_0$,
which will be shown in Figs. \ref{ene_err_positive} and \ref{ene_err_negative}.
We also calculate standard deviation of the energies which will be
shown by the error bar in the figure. The standard deviation is defined by
\begin{equation}
  \sigma^{J\pi}_n=\sqrt{\frac{1}{N_s}\sum_{i=1}^{N_s}(E^{J\pi}_{n,i}
  -\overline E^{J\pi}_n)^2}.
  \label{e:sd}
\end{equation}
The average values and the standard deviations for the transition strengths
and radii are evaluated in the same way.

\subsection{Energy levels}\label{s:el}

\begin{figure}[Htb]
  \includegraphics[width=82mm]{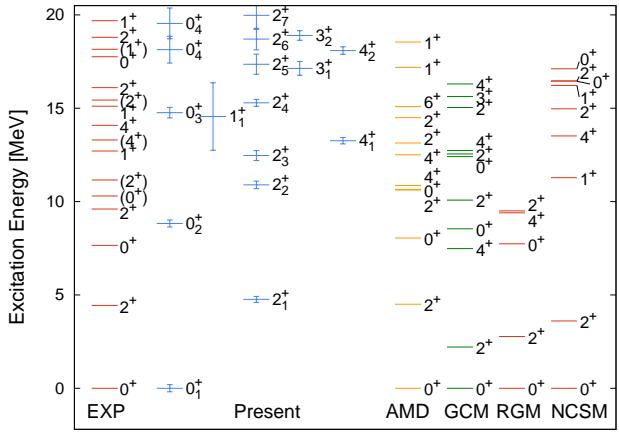}
  \caption{Excitation energies of positive parity are shown. 
    The energies are obtained by averaging over ten configurations. 
    The standard deviations of the energies are also shown by error
    bars. We also show the results of AMD \cite{Kanad2007}, 
    GCM \cite{EUega1979}, RGM \cite{MKamimura1981} and 
    NCSM \cite{PNavratil2003}. See the text for details.}
  \label{ene_err_positive}
\end{figure}
\begin{figure}[Htb]
  \includegraphics[width=82mm]{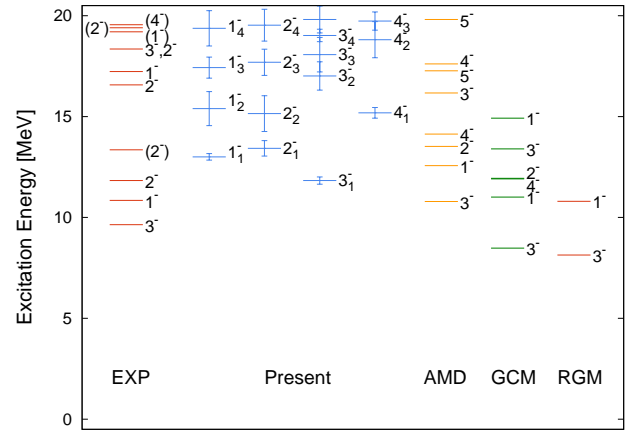}
  \caption{Excitation energies of negative parity states. 
  The same explanation as that in Fig. \ref{ene_err_positive} applies 
  to others.}
  \label{ene_err_negative}
\end{figure}

We show calculated excitation spectra of even and odd parities
in Figs.~\ref{ene_err_positive} and \ref{ene_err_negative}, respectively. 
In the figures, energies averaged over ten sets are shown with
error bars as the standard deviation. Our calculated results are 
compared with measurements
\cite{Ajzenberg1990,WRZimmerman2011,PhysRevC.84.054308,PhysRevC.86.034320,PhysRevC.83.034314} 
and other theories, AMD \cite{Kanad2007}, GCM \cite{EUega1979}, 
RGM \cite{MKamimura1981}, and NCSM \cite{PNavratil2003}. 

In the Skyrme-HF calculation,
the binding energy of $^{12}$C is 90.6 MeV, in reasonable
agreement with the measured value, 92.2 MeV.
In our configuration mixing calculation, the correlation energy
is $4.7 \pm 0.2$ MeV. The ground-state energy 
including the correlation is $95.3 \pm 0.2$MeV, slightly lower 
than the measured value.

Calculated excitation energies of the ground-state rotational
band are in good agreement with measurements. 
In the nice reproduction of the rotational energy levels,
the configuration mixing is essential since the ground state solution
in the HF calculation is spherical for $^{12}$C with the
SLy4 interaction. The excitation 
energy of $2_1^+$ state is well reproduced by the present calculation,
AMD, and NCSM.
However, microscopic $\alpha$ cluster models (GCM and RGM)
provide too low excitation energies.
The former models (present, AMD, and NCSM)
take into account the spin-orbit interaction, while it is not included 
in the latter (GCM and RGM) in which existence of the three $\alpha$ 
clusters are assumed.
This suggests that a proper inclusion of the spin-orbit interaction
is important for the good description of the ground rotational band.

The $0_2^+$ state, known as the Hoyle state,
has been attracting much attention 
recently since it has been shown that this state may be 
understood as the Bose condensed state of three $\alpha$ 
particles \cite{YFunaki2003}. 
Our calculation gives a reasonable description for this state,
although the excitation energy is slightly overestimated by about 1 MeV.
Properties of this state will be discussed in the following subsections.
Although recent ab-initio approaches have been successful for the
ground-state rotational band, a satisfactory description for 
the $0_2^+$ state has not yet been made. For example, the NCSM cannot
describe this state adequately \cite{PNavratil2003}. 
Recently, attempts of ab-initio description for this state have been
undertaken by several groups. A nuclear lattice calculation for this state
has been reported in Ref. \cite{PhysRevLett.106.192501}.

Recently, a new $2^+$ state has been found experimentally at $9.6\pm0.1$ MeV 
excitation energy with a width of $0.6\pm0.1$ MeV \cite{MFree2009,WRZimmerman2011}
This state was interpreted as the excited state built on the 
$0_2^+$ state. In our calculation, two $2^+$ states, $2_2^+$ and
$2_3^+$, appear just above the $0_2^+$ state. However, as we discuss in
Sec.~\ref{s:trans},
these two states, $2_2^+$ and $2_3^+$, in the present calculation,
seem not to correspond 
to rotationally excited state of the Hoyle state.

In Fig.~\ref{ene_err_positive}, three states, $0_3^+$, $2_4^+$, 
and $4_2^+$, follow a rotational energy sequence. Small standard 
deviations of the energies of these states indicate the reliability of the 
calculation. As will be discussed in Sec.~\ref{s:trans},
these states are connected by strong $B(E2)$ transitions.
In Sec.~\ref{s:ana_wf},
we will show that this band corresponds to a three-$\alpha$ linear-chain state.

For the negative parity states, we have obtained solid results only for the
lowest energy state for each $J^{\pi}$ sector (Sec. \ref{s:conv_el}).
Our calculation reproduces the measured order of the three states, 
$3_1^-$, $1_1^-$ and $2_1^-$. However, the excitation energies 
are higher than the measurements by 2-3 MeV.

We would like to stress that our calculation includes no empirical
parameter specific to the system, $^{12}$C nucleus. We employ the
SLy4 parameter set which is determined to reproduce nuclear properties
of whole mass region. This is in contrast to cluster model calculations
where nuclear force parameters are often adjusted for respective systems.
We also do not employ any effective charges in
evaluating the transition matrix elements shown below.

\begin{figure}[Htb]
  \includegraphics[width=82mm]{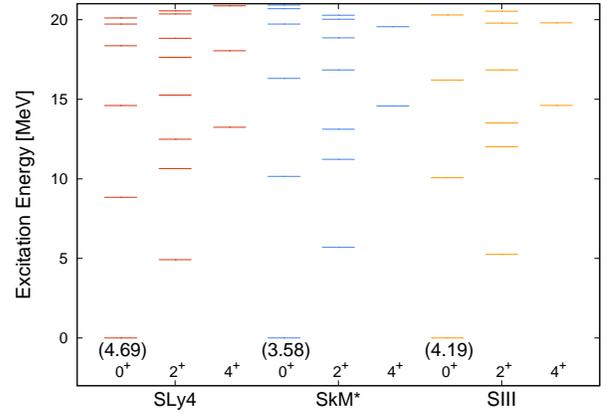}
  \caption{Energy levels of positive parity employing different parameter sets of
    Skyrme force, SLy4, SkM* and SIII. The number in parentheses is the
    correlation energy in the ground state, $E_{\rm HF}-E_{\rm gs}$,
    in unit of MeV.
    In the calculation, the same set of SDs is employed.}
  \label{f:ene_dif_Skyrme}
\end{figure}

Finally we mention how the calculated energy levels depend on the 
choice of the Skyrme interaction.
In Fig.~\ref{f:ene_dif_Skyrme}, we show the excitation energies of 
positive parity states with different parameter sets of the Skyrme interaction, 
SLy4, SkM*, and SIII. The same set of SDs 
(No. 1 in Fig.~\ref{f:ene_dif}) is employed in all calculations. 
The correlation energies in the ground state are shown as well 
inside the parentheses.
The comparison shows that basic features of the spectra do not
depend much on the choice of the Skyrme parameters. For example,
the ground rotational band is described reasonably by all three
parameter sets. The position of $0_2^+$ state does not change much.
There appear rotational band in three calculations starting 
with $0_3^+$ state. We thus conclude that the excitation 
energies are not sensitive to the choice of the Skyrme parameters 
for almost all the states below 15 MeV.

\subsection{Transition strength}\label{s:trans}

\begin{table*}[Htb]
  \begin{center}
    \begin{tabular}{c|ccccccc}
      \hline\hline Transitions & Exp & Cal & AMD & GCM & RGM & NCSM & THSR\\
      \hline
      $B(E2;2_1^+ \rightarrow 0_1^+)$ & 7.6$\pm$0.4 & 8.6 $\pm$0.2  & 8.5     & 8.0 & 9.3 & 4.146　& 9.06 \\ 
      $B(E2;4_1^+ \rightarrow 2_1^+)$ &             & 13.4$\pm$0.5  & 16      &     &     &       & 10.73 \\ 
      $B(E2;0_2^+ \rightarrow 2_1^+)$ & 13$\pm$2    & 13.6$\pm$1.2  & 25.5    & 3.5 & 5.5 &       & 4.71 \\ 
      $B(E2;2_2^+ \rightarrow 0_2^+)$ &             & 0.17$\pm$0.23 &         &     &     & \\
      $B(E2;2_3^+ \rightarrow 0_2^+)$ &             & 5.9$\pm$0.7   &         &     &     & \\ 
      $B(E2;2_4^+ \rightarrow 0_2^+)$ &             & 10$\pm$1      & 100$^*$ &     &     &       & 391 \\
      $B(E2;2_4^+ \rightarrow 0_3^+)$ &             & 91$\pm$13     & 310$^*$ &     &     & \\
      $B(E2;4_2^+ \rightarrow 2_4^+)$ &             & 131$\pm$22    & 600$^*$ &     &     & \\
      $B(E3;3_1^- \rightarrow 0_1^+)$ & 107$\pm$14  & 77$\pm$4      &         & 99  & 124 \\
      $M(E0;0_1^+ \rightarrow 0_2^+)$ & 5.4$\pm$0.2 & 4.5$\pm$0.2   & 6.7     & 6.6 & 6.7 &       & 6.50 \\
      \hline 
    \end{tabular}
  \end{center}
  \caption{$B(E2)$, $B(E3)$ and $M(E0)$ values 
  of $^{12}$C in units of $e^2{\rm fm}^4$, $e^2{\rm fm}^6$ and $e{\rm fm}^2$ 
  respectively. 
  Experimental and calculated values are shown in the first and second
  column, respectively.
  For comparison, results of the AMD\cite{Kanad2007}, 
  GCM\cite{EUega1979}, RGM\cite{MKamimura1981},  
  NCSM\cite{PNavratil2003}, and THSR are shown.
  Values in THSR are calculated with the same model as Ref.~\cite{YFunaki2003}.  
  The values indicated by * correspond to 
  $B(E2;2_2^+ \rightarrow 0_2^+)$, $B(E2;2_2^+ \rightarrow 0_3^+)$ 
  and $B(E2;4_2^+ \rightarrow 2_2^+)$ in Ref.~\cite{Kanad2007}.
  See text for details.}
  \label{e2m0}
\end{table*}

Calculated $B(E2)$, $B(E3)$, and $M(E0)$ values,
the average values and the standard deviations, are shown
in Table~\ref{e2m0}.
In our calculated values, we do not employ any effective charges. 
The $B(E2)$ transition strength between $2_1^+$ and $0_1^+$ states is 
well reproduced by our calculation. It is also consistent with 
results of other theories. The standard deviation is small, 
about 3\%, indicating that our calculated value is well converged.

The $B(E2)$ transition between $0_2^+$ and $2_1^+$
is calculated as $13.6 \pm 1.2 e^2{\rm fm}^4$, which is in excellent 
agreement with the measured value, $13 \pm 2 e^2{\rm fm}^4$. Other 
theories fail to reproduce the rate.
In Ref.~\cite{Kanad2007}, it is argued that this transition strength 
is sensitive to the $\alpha$-breaking effect. A good reproduction 
of this transition strength by our calculation indicates that our
calculation reasonably takes account of the $\alpha$-cluster
components in the states.

As mentioned in Sec.~\ref{s:el}, there appear two $2^+$ states,
$2_2^+$ and $2_3^+$, above the $0_2^+$ state in our calculation.
These states might correspond to the $2^+$ state 
at 9.6 MeV which was discovered recently \cite{MFree2009,WRZimmerman2011}.
It was suggested to be a candidate of rotationally 
excited state built on the $0_2^+$ state.
In the present calculation, however, the $B(E2)$ transition strength 
between $0_2^+$ and $2_2^+$ states is small as seen in 
Table~\ref{e2m0}.
The rate between $0_2^+$ and $2_3^+$ states is also not very large.
The $B(E2)$ rate from $0_2^+$ state is the largest for $2_4^+$ state
which is regarded as rotationally excited state built on the $0_3^+$ state, 
as will be mentioned below.
These observations suggest that the $2_2^+$ and 
$2_3^+$ states in the present calculation do not correspond to a rotationally
excited state on the $0_2^+$ state.
 
As we discussed in Sec.~\ref{s:el}, the states of $0_3^+$,
$2_4^+$, and $4_2^+$ follow the rotational energy sequence. 
The calculated transition strengths of $B(E2;2_4^+ \rightarrow 0_3^+)$
and $B(E2;4_2^+ \rightarrow 2_4^+)$ are very large.
These results strongly support that these states indeed constitute 
a rotational band. In Sec.~\ref{s:ana_wf}, we will see that this band 
is dominated by the three-$\alpha$ linear-chain structure.

In the AMD calculation \cite{Kanad2007}, intense $B(E2)$ values are
reported in the transitions among $0_3^+$, $2_2^+$, and $4_2^+$ states. 
Since the states corresponding to $2_2^+$ and $2_3^+$
in our calculation seem not to  be present in the AMD calculation,
we put these $B(E2)$ values by AMD at the places of $B(E2; 2_4^+ \rightarrow
0_3^+)$ and $B(E2; 4_2^+ \rightarrow 2_4^+)$ in Table~\ref{e2m0}.
In Ref. \cite{Kanad2007}, these states are considered as
the three-$\alpha$ linear-chain states.
The large $B(E2)$ values are
qualitatively consistent with our result, though the absolute magnitudes 
of the transition strengths are much smaller in the present calculation.

For negative-parity states, experimental data for
$B(E3; 3_1^- \rightarrow 0_1^+)$
are available.
The present calculation gives
$77 \pm 4 e^2{\rm fm}^6$, which is slightly 
smaller than the measured value, $107 \pm 14 e^2 {\rm fm}^6$.

Finally, we discuss the $M(E0)$ transition strength between 
$0_2^+$ and $0_1^+$ states. In the studies by cluster models,
it has been argued that the magnitude of this transition strength
reflects the spatial extension of the $0_2^+$ state \cite{TYamada2008}. 
Our calculated value, $4.5 \pm 0.2 e {\rm fm}^2$, is slightly smaller than
the measured value, $5.4 \pm 0.2 e {\rm fm}^2$.
In contrast, microscopic cluster models and AMD have
reported an opposite feature, slightly larger values, $6.6 - 6.7$ $e$ fm$^2$,
than measurement\cite{EUega1979,MKamimura1981,Kanad2007}. 
Experimentally measured value, 
$5.4 \pm 0.2 e {\rm fm}^2$ \cite{Ajzenberg1990}, is located between our
result and those of the other calculations.

\subsection{Radii}
We next examine root-mean-square (rms) radii of the ground and excited states.
Since our wave function does not allow an exact separation of 
the center-of-mass motion from the internal one, we estimate an approximate
correction for the radius due to the center-of-mass motion, 
and subtract it from the calculated values. 
We assume a harmonic oscillator motion for the center-of-mass
with the oscillator constant given by $\hbar\omega = 41 A^{-1/3}=17.9$ MeV.
The value of the correction in this model is 
estimated to be 0.07 fm in the harmonic oscillator shell model.
The calculated radii after the correction are shown in Table~\ref{f:radii}. 

Our calculated value in the ground state is $2.52 \pm 0.01$ fm.
This value is somewhat larger than the measured value, $2.31 \pm 0.02$ fm.
In the HF calculation, the radius is given by
$2.44$ fm. Our configuration mixing calculation, therefore,
slightly increases the radius.
Comparing with other theories, our value is larger than those of
GCM and FMD, and is comparable to the value of AMD.

For the $2_1^+$ state, our calculated radius is slightly larger
than that of the ground state. Other theories 
report almost the same or slightly larger radius for this state.

For the $0_2^+$ state, we find a significant difference between
the present calculation and the others.
Our calculated radius is $2.73 \pm 0.02$ fm,
which is larger than the radius in the ground state.
However, this is much smaller than the other calculations which
give more than 3 fm \cite{Kanad2007,EUega1979,MKamimura1981,MChernykh2007,YFunaki2003}. 
In the recent AMD+GCM calculation \cite{Suhara2010},
the radius of 2.9 fm was reported, similar to ours.
It has been found that the radius of the $0_2^+$ state 
is quite sensitive to the spin-orbit interaction used in the AMD 
calculation \cite{Suhara-Enyo}.
The radius of the $0_2^+$ state decreases
as the strength of the spin-orbit interaction increases.
This dependence is understood as follows \cite{Suhara-Enyo}.
If the strength of the spin-orbit interaction is weak, the ground state 
wave function contains substantial amount of 
the three-$\alpha$-cluster component.
Then, the  $0_2^+$ wave function, which is dominated by dilute
three-$\alpha$ components, spatially expands to ensure the orthogonalization 
to the ground state. As the spin-orbit interaction increases, 
the three-$\alpha$ component decreases in the ground state, which allows
$0_2^+$ wave function to include more compact three-$\alpha$ structure. 
This change results in decrease of the radius in $0_2^+$ state.
This mechanism may explain the discrepancy in the $0_2^+$ radius
between our calculation and other theories.
It should be noted again that our calculated value for 
the $M(E0)$ transition strength is smaller than the calculated values by
other theories.
We also note that an indirect measurement of radius
for the $0_2^+$ state using diffraction inelastic scattering 
\cite{PhysRevC.80.054603} was reported, giving $2.89 \pm 0.04$ fm. 

For the $0_3^+$ state, our calculated radius is $3.20 \pm 0.05$ fm,
which is much larger than the radii of $0_1^+$ and $0_2^+$ states. 
This is again smaller than those by other models listed in
Table.\ref{f:radii}, while it is similar to the value (3.26 fm)
in Ref.~\cite{Suhara2010}.

\begin{table}[Htb]
  \begin{tabular}{c|cccccccc}
  \hline\hline
  $J^{\pi}$ &EXP & present       & AMD  & FMD  & GCM  & RGM  & THSR \\
  \hline
  $0_1^+$&2.31(2)& 2.52$\pm$0.01 & 2.53 & 2.39 & 2.40 & 2.40 & 2.39 \\
  $0_2^+$&       & 2.73$\pm$0.02 & 3.27 & 3.38 & 3.40 & 3.47 & 3.80\\
  $0_3^+$&       & 3.20$\pm$0.05 & 3.98 & 4.62 & 3.52 &      &\\
  $2_1^+$&       & 2.60$\pm$0.01 & 2.66 & 2.50 & 2.36 & 2.38 & 2.36\\
  \hline 
  \end{tabular}
  \caption{Mass rms radii of the ground and excited states of $^{12}$C. 
   The experimental data is taken from Ref. \cite{AOzawa2001}. For comparison, we show 
   the results of AMD \cite{Kanad2007}, FMD \cite{MChernykh2007}, GCM \cite{EUega1979}
   , RGM \cite{MKamimura1981} and THSR.}
  \label{f:radii}
\end{table}

\subsection{Charge form factors}

\begin{figure*}[Htb]
  \begin{center}
    \includegraphics[width=15.2cm]{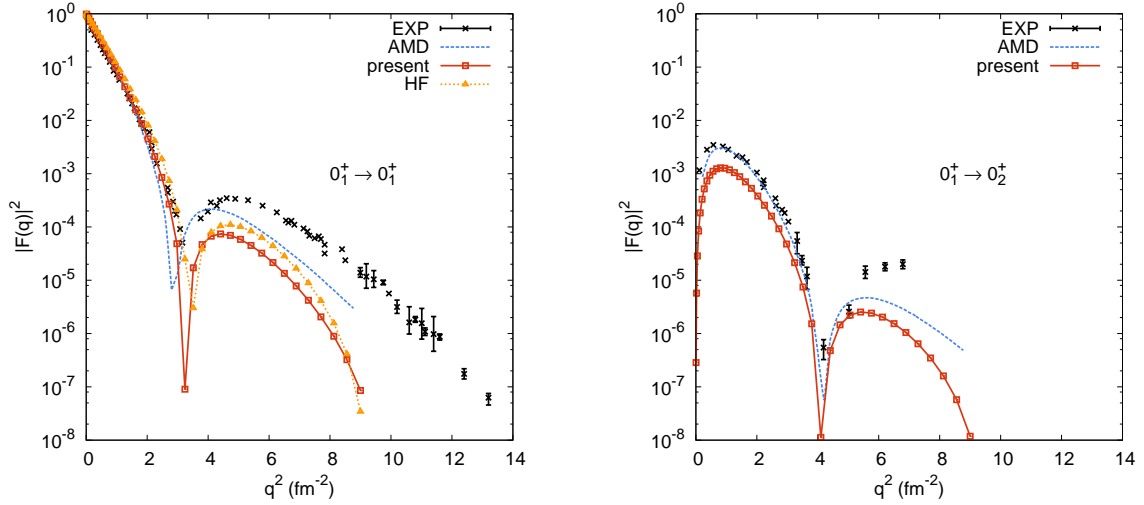}
  \end{center}
  \caption{Squared elastic form factor for the ground state (left) and 
  $0_1^+\rightarrow 0_2^+$ inelastic form factor (right) are shown. Our
  calculated results are compared with the HF calculation with
  a single SD, measurements  
  \cite{Sick1970631,Nakad1971,Nakad1971_e,Streh1968}
  and the AMD calculation \cite{Kanad2007_e}.
  Here, the result of the first set of SDs in Fig.~\ref{f:ene_dif} is used.
}
  \label{cff010}
\end{figure*}

A charge form factor from the initial state $\big|J_i, M_i\big>$ to the 
final state $\big|J_f, M_f\big>$ is defined as follows,
\begin{eqnarray}
 &|F_{J_i\rightarrow J_f}(q^2)|^2&=\frac{1}{Z}\frac{1}{2J_i+1}\nonumber\\
&& \times\sum_{M_iM_j}\big|\big<J_f M_f|\sum_k\frac{1+\tau_z(k)}{2}e^{i\vec q\cdot \vec r_k}|J_iM_i\big>\big|^2\nonumber\\
&& \times F_p^2(q^2) \times F_{\rm cm}^2(q^2), \label{e:cff}
\end{eqnarray}
where $Z$ is the proton number and $\vec q$ is the transferred momentum. 
$F_p(q^2)$ is a correction factor for the proton size for which we employ 
$F_p(q^2)={\rm exp}(-a_p^2q^2/6)$ with $a_p = 0.831$ fm. 
To correct the center-of-mass motion, we simply assume that
the center-of-mass motion is separated and described by the
harmonic oscillator wave function of the same oscillator
constant, $\hbar\omega=41 A^{-1/3}$ MeV, for both initial and
final states. Thus, this leads to
$F_{\rm cm}^2(q^2)=\exp(q^2b^2/2A)$ with $b=1.66$ fm.

In Fig. \ref{cff010}, we show charge form factors for the
elastic (left) and inelastic $0_1^+ \rightarrow 0_2^+$ (right) 
processes. Red solid curves show our results, blue dashed curves
show the results of AMD calculation \cite{Kanad2007_e}, and crosses with
error bars show experimental results \cite{Sick1970631,Nakad1971,Nakad1971_e,Streh1968}.
For the elastic form factor, we also show that of 
Skyrme HF solution in the ground state.

In the small momentum transfer region $q^2<2$ fm$^{-2}$, 
the elastic form factor is well reproduced by the calculation.
For $q^2>2$ fm$^{-2}$, our calculation underestimates the
form factor, though position of the dip at around 3 fm$^{-2}$
is reproduced well.
The inelastic form factor for $0_1^+ \rightarrow 0_2^+$ transition
is underestimated for a whole momentum transfer region.
The position of the dip at around 4 fm$^{-2}$ is reproduced well.

We show results by the AMD calculation. They are in better 
agreement with measured values, although the dip position
in the elastic form factor is located at somewhat smaller $q^2$ value.
Microscopic cluster calculations also reproduce the form
factors well \cite{EUega1979,MKamimura1981}. 

The underestimation of the elastic form factor at large $q^2$ 
value indicates that the density in our calculation lacks 
high momentum component.
Since the HF solution gives a better description for the form 
factor at high momentum, the superposition of a number of 
Slater determinants turns out to increase 
the diffuseness in the nuclear surface,
making the density distribution function $\rho(r)$ smoother.
As for the underestimation in the inelastic form factor 
of $0_1^+ \rightarrow 0_2^+$ transition, a possible reason
is the difference in the character of the wave functions
between the ground and $0_2^+$ states. As we discussed in the
radii, a rather small radius of $0_2^+$ state in our calculation 
may indicate a small fraction of three-alpha component in the 
ground state. The inelastic form factor may be reduced if the
correlation structures are different between two states.
It has been argued that the magnitude of this
form factor at small $q^2$ is quite sensitive to the radius 
of the $0_2^+$ state \cite{YFunaki2006_2}: the magnitude
of the form factor at small $q^2$ reduces as the radius of 
the $0_2^+$ state increases. Our result here is opposite,
however. The radius of $0_2^+$ state in our calculation is
smaller than those by cluster models, and the magnitude
of the inelastic form factor is also small.

\subsection{Analysis for wave functions}
\label{s:ana_wf}

In order to clarify what kinds of correlations are included
in the wave function after configuration mixing, $\Psi_n^{J\pi}$,
we calculate the overlap between the energy eigenstate and the
projected single SD state,
\begin{eqnarray}
P_n^{J\pi,iK} &=&
 \frac{\Big| \big< \Phi_i \big|\hat P^{J\dagger}_{MK}\hat P^\pi|\Psi^{J\pi}_{n}\big>\Big|^2}
   { \big| \big< \Phi_i \big| \hat P^J_{KK}\hat P^\pi \big|\Phi_i\big>\big| }\nonumber\\
     &=& \Bigg| \sum_{j\nu}f^{J\pi,n}_{j\nu}\frac{1}{\sqrt{e^{J\pi}_{j\nu}}}\sum_{K'}v^{J\pi,j\nu}_{K'} \frac{\big<\Phi_i\big|\hat P^J_{KK'}\hat P^\pi\big|\Phi_j\big>}{\sqrt { \big< \Phi_i \big| \hat P^J_{KK}\hat P^\pi \big|\Phi_i\big>}}\Bigg|^2,\nonumber\\
\label{e:overlap}
\end{eqnarray}
and find the SDs which have large overlap values with $\Psi_n^{J\pi}$.
We show density distributions of the SDs to visualize the
correlations included.

In the following, we use the sequential number of the SDs which
we assigned in Sec.~\ref{s:Generation_SD}, using the result
of the first set of SDs in Fig.~\ref{f:ene_dif}.
We also show the $K$ quantum number of the SD
and the value of the overlap, $P_n^{J\pi,iK}$, in Eq.~(\ref{e:overlap}).

\subsubsection{The ground rotational band}
 
\begin{table}[Htb]
  \begin{tabular}{cc|c|cc|c|cc|c}
    \hline\hline
    \multicolumn{3}{c|}{$0^+_1$} & \multicolumn{3}{c|}{$2^+_1$}
    & \multicolumn{3}{c}{$4^+_1$}\\
    \hline
    SD & $K^{\pi}$ & \% & SD & $K^{\pi}$ & \% & SD & $K^{\pi}$ & \% \\ 
    \hline
    15 & $0^+$ & 90.38 & 4  & $0^+$ & 89.36 & 4  & $0^+$ & 88.60 \\
    7  & $0^+$ & 87.44 & 15 & $0^+$ & 88.51 & 15 & $0^+$ & 81.02 \\
    8  & $0^+$ & 84.78 & 29 & $0^+$ & 82.44 & 29 & $0^+$ & 76.86 \\
    31 & $0^+$ & 84.75 & 2  & $0^+$ & 76.47 & 7  & $0^+$ & 76.60 \\
    2  & $0^+$ & 81.69 & 7  & $0^+$ & 75.21 & 29 & $1^+$ & 72.63 \\
    42 & $0^+$ & 80.10 & 48 & $0^+$ & 72.63 & 15 & $1^+$ & 70.60 \\
    24 & $0^+$ & 80.05 & 47 & $0^+$ & 65.76 & 3  & $0^+$ & 70.46 \\
    4  & $0^+$ & 79.17 & 8  & $0^+$ & 64.22 & 47 & $0^+$ & 70.28 \\
    16 & $0^+$ & 78.71 & 10 & $1^+$ & 63.85 & 48 & $0^+$ & 69.85 \\
    35 & $0^+$ & 77.30 & 44 & $1^+$ & 63.63 & 2  & $0^+$ & 69.66 \\
    \hline
  \end{tabular}
  \caption{The sequential number, the $K$-value, and the squared 
  overlap value are shown for the SDs which dominate in the wave 
  function of the ground rotational band.}
  \label{over_ground}
\end{table}

In Table~\ref{over_ground}, we show the sequential number 
of the SDs which have large overlap values with the wave function of
the ground rotational band, $0_1^+$, $2_1^+$, and $4_1^+$.
The overlap values $P_n^{J\pi,iK}$ defined by Eq.~(\ref{e:overlap})
and $K$ values are shown as well.
Since the SDs are non-orthogonal, the sum of the overlap
values is not equal to but much larger than unity. 

\begin{figure}[Htb]
  \begin{center}
    \includegraphics[width=8.0cm]{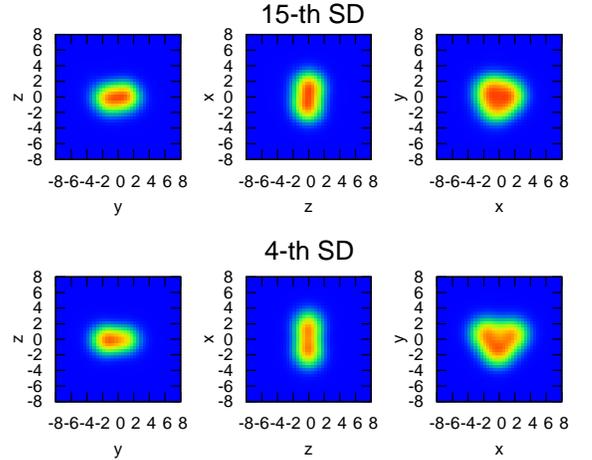}
  \end{center}
  \caption{Contour plots of the density distributions of the 15th and the 4th SDs, 
  which are the major components of the ground state rotational band.}
  \label{dens_ground}
\end{figure}

In the ground state $0_1^+$,  the 15th SD has the largest overlap, 
showing 90\% for the overlap value. In $2_1^+$ and $4_1^+$ states, 
the 4th SD is the largest component and the 15th SD is the second largest.
To illustrate the nuclear shape of these two SDs, we show 
contour plots of density distributions of the SDs in the {\it yz}, {\it zx}, 
and {\it xy} planes in Fig.~\ref{dens_ground}. As seen from the figure, 
they both show oblate deformed shapes.

The self-consistent HF solution is assigned to the first SD (number 1).
We should note that it does not appear in the 
top ten components of the ground state. Its overlap value
with $0_1^+$ is about 70\%.
For $^{12}$C, the self-consistent HF solution with the SLy4
interaction is spherical with $p_{3/2}$ closed shell configuration. 
The spherical solution cannot describe the rotational band observed 
in the measurement.
As shown in Fig.~\ref{ene_err_positive} and Table~\ref{e2m0}, 
our calculation accurately reproduces the energy levels and the $B(E2)$ 
transitions among the states of the ground rotational band. This good 
reproduction is achieved by a superposition of SDs of deformed
shapes.

\subsubsection{Negative-parity states}

In Table~\ref{over_negative}, we show sequential numbers of the
SD which have large overlap values with the wave function of
the negative parity states, $3_1^-$, $1_1^-$, and $2_1^-$.
We find the 4th SD, which appears in the ground rotational band, 
also dominates in the negative parity states. Other SDs which dominate
in the negative-parity states are 21, 29, and 47. 

\begin{figure}[Htb]
  \begin{center}
    \includegraphics[width=8.0cm]{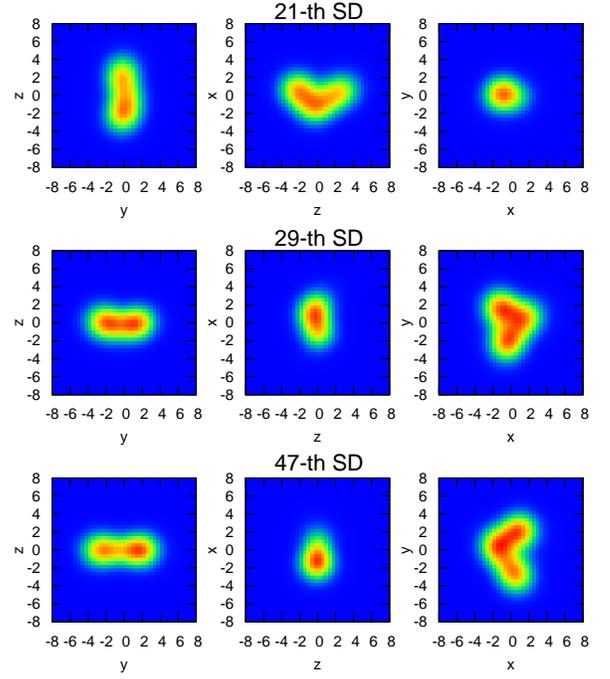}
  \end{center}
  \caption{Contour plots of the density distributions of the 21th, 29th, and 47th SDs,
               which are major components of the negative parity states, 
                $1^-_1$, $2^-_1$, $3^-_1$ and $4^-_1$.}
  \label{dens_negative}
\end{figure}

We show density distributions of these three SDs in
Fig.~\ref{dens_negative}. 
All of these SDs have similar oblate shapes with three-$\alpha$-like
structure. 
Close look at the densities reveals that
the 4th and the 29th SDs have a compact configuration,
while the 21th and the 47th SDs show spatially more extended 
three-$\alpha$ configurations forming an obtuse-angled triangle. 

\begin{table}[Htb]
  \begin{tabular}{cc|c|cc|c|cc|c|cc|c}
    \hline\hline
    \multicolumn{3}{c|}{$1^-_1$} & \multicolumn{3}{c|}{$2^-_1$} 
    & \multicolumn{3}{c|}{$3^-_1$} & \multicolumn{3}{c}{$4^-_1$}\\
    \hline
    SD  & $K^{\pi}$ & \% & SD  & $K^{\pi}$ & \% & SD  & $K^{\pi}$ & \% & SD  & $K^{\pi}$ & \%  \\ 
    \hline
    4  & $1^-$ & 76.80 & 21 & $1^-$ & 76.50 & 29 & $3^-$ & 81.34 & 29 & $3^-$ & 79.34 \\
    47 & $1^-$ & 76.35 & 4  & $1^-$ & 75.36 & 4  & $3^-$ & 81.10 & 47 & $3^-$ & 77.26 \\
    18 & $1^-$ & 75.28 & 47 & $1^-$ & 71.96 & 47 & $3^-$ & 76.17 & 4  & $3^-$ & 76.43 \\
    21 & $1^-$ & 75.26 & 22 & $1^-$ & 70.24 & 15 & $3^-$ & 66.77 & 25 & $2^-$ & 66.05 \\
    22 & $1^-$ & 74.03 & 18 & $1^-$ & 69.73 & 21 & $3^-$ & 63.91 & 41 & $3^-$ & 64.37 \\
    29 & $1^-$ & 68.00 & 29 & $1^-$ & 69.72 & 3  & $1^-$ & 63.63 & 3  & $3^-$ & 60.82 \\
    11 & $1^-$ & 67.24 & 11 & $1^-$ & 61.80 & 48 & $3^-$ & 61.51 & 5  & $2^-$ & 60.20 \\
    25 & $1^-$ & 59.97 & 48 & $1^-$ & 55.10 & 9  & $3^-$ & 60.52 & 9  & $3^-$ & 58.79 \\
    46 & $1^-$ & 57.62 & 46 & $1^-$ & 54.17 & 41 & $1^-$ & 55.94 & 48 & $3^-$ & 56.95 \\
    33 & $1^-$ & 57.27 & 33 & $1^-$ & 53.68 & 33 & $3^- $ & 54.83 & 21 & $1^-$ & 55.25 \\
    \hline
  \end{tabular}
  \caption{The sequential number, the $K$-value, and the squared 
  overlap value are shown for the SDs which dominate in the wave 
  function of the negative parity states $1_1^-$, $2_1^-$, $3_1^-$ and $4_1^-$.}
  \label{over_negative}
\end{table}

\subsubsection{$0_2^+$, $2_2^+$ and $2_3^+$ states}

In Table~\ref{over_Hoyle}, we show sequential numbers of the
SDs which have large overlap values with the wave function of
the states, $0_2^+$, $2_2^+$ and $2_3^+$.

We first examine the Hoyle state, $0_2^+$.
Compared with the cases of the ground rotational band and the negative-parity 
states, the maximum value of the overlap is rather small, less than 50\%.
This indicates that the superposition of a large number of SDs is essential to 
describe the $0_2^+$ state. This is consistent with the cluster-model
calculations \cite{EUega1979,MKamimura1981}
and the picture of 
the $\alpha$ condensed state for the $0_2^+$ state \cite{YFunaki2003}.

\begin{table}[Htb]
  \begin{tabular}{cc|c|cc|c|cc|c}
    \hline\hline
    \multicolumn{3}{c|}{$0^+_2$} & \multicolumn{3}{c|}{$2^+_2$} & \multicolumn{3}{c}{$2^+_3$}\\
    \hline
    SD  & $K^{\pi}$ & \% & SD  & $K^{\pi}$ & \% & SD  & $K^{\pi}$ & \% \\ 
    \hline
    9  & $0^+$ & 46.26  & 16 & $2^+$ & 74.07 & 15 & $2^+$ & 50.05 \\
    28 & $0^+$ & 44.66 & 35 & $0^+$ & 58.22 & 7  & $2^+$ & 40.80 \\
    3  & $0^+$ & 41.21 & 43 & $0^+$ & 56.18 & 8  & $2^+$ & 40.29 \\
    5  & $0^+$ & 39.44 & 42 & $0^+$ & 56.00 & 31 & $0^+$ & 39.33 \\
    33 & $0^+$ & 38.91 & 31 & $2^+$ & 55.69 & 16 & $0^+$ & 31.33 \\
    11 & $0^+$ & 35.96 &  7 & $1^+$ & 53.55 & 32 & $0^+$ & 22.94 \\
    47 & $0^+$ & 35.25 & 32 & $2^+$ &51.61  & 43 & $2^+$ & 21.80 \\
    26 & $0^+$ & 33.97 & 49 & $0^+$ & 41.33 & 10 & $2^+$ & 20.14 \\
    45 & $0^+$ & 33.27 & 36 & $2^+$ & 41.02 & 35 & $2^+$ & 19.20 \\
    41 & $0^+$ & 32.30 & 43 & $1^+$ & 36.06 & 4  & $2^+$ & 18.81 \\
    \hline
  \end{tabular}
  \caption{The sequential number, the $K$-value, and the squared 
  overlap value are shown for the SDs which dominate in the wave 
  function of the $0_2^+$m $2_2^+$ and $2_3^+$.}
  \label{over_Hoyle}
\end{table}

\begin{figure}[Htb]
  \begin{center}
     \includegraphics[width=8.0cm]{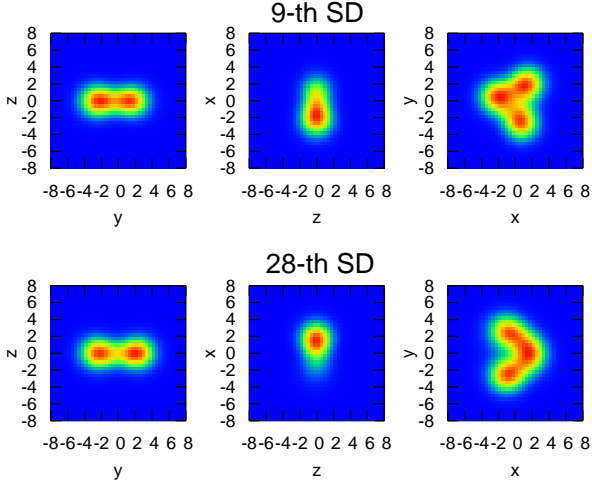}
  \end{center}
  \caption{ Contour plots of the density distributions of the 9th and the 28th SDs,
               which are major components in the $0_2^+$ state.}
  \label{dens_Hoyle}
\end{figure}

We show in Fig. \ref{dens_Hoyle} the density distributions of 
the  SDs which have the largest and next largest overlaps with
the $0_2^+$ state, namely, the 9th and the 28th SDs.
These SDs have a well developed
cluster structures of three $\alpha$-particles.

Regarding the $2_2^+$ and the $2_3^+$ states, we find that a number of
configurations contribute to these states, as in the case of $0_2^+$ state.
The SDs in $2_2^+$ and $2_3^+$ are more or less similar.
However, they are very different from those in the $0_2^+$ state.
 This is consistent with our observation that the $B(E2)$
transition strengths between $0_2^+$ and $2_2^+$ states, and between $0_2^+$
and $2_3^+$ states are rather small (see Sec. \ref{s:trans}). 

\subsubsection{Linear-chain states}

\begin{table}[Htb]
  \begin{tabular}{cc|c|cc|c|cc|c}
    \hline\hline
    \multicolumn{3}{c|}{$0^+_3$} & \multicolumn{3}{c|}{$2^+_4$} & \multicolumn{3}{c}{$4^+_2$}\\
    \hline
    SD  & $K^{\pi}$ & \% & SD  & $K^{\pi}$ & \% & SD  & $K^{\pi}$ & \%  \\ 
    \hline
    30 & $0^+$ & 70.13 & 40 & $0^+$ & 78.31 & 40 & $0^+$ & 75.26 \\
    40 & $0^+$ & 66.70 & 30 & $0^+$ & 72.35 & 30 & $0^+$ & 75.18 \\
    19 & $0^+$ & 65.11 & 19 & $0^+$ & 71.33 & 18 & $0^+$ & 65.74 \\
    20 & $0^+$ & 41.88 & 18 & $0^+$ & 67.69 & 19 & $0^+$ & 62.44 \\
    23 & $0^+$ & 38.47 & 11 & $0^+$ & 59.29 & 34 & $0^+$ & 43.14 \\
    18 & $0^+$ & 38.02 & 23 & $0^+$ & 57.94 & 20 & $0^+$ & 43.05 \\
    14 & $0^+$ & 37.56 & 12 & $0^+$ & 47.82 & 11 & $1^+$ & 41.81 \\
    12 & $0^+$ & 37.23 & 34 & $0^+$ & 47.07 & 23 & $0^+$ & 41.19 \\
    11 & $0^+$ & 25.46 & 22 & $0^+$ & 39.24 & 14 & $0^+$ & 40.48 \\
    22 & $0^+$ & 16.63 & 20 & $0^+$ & 39.15 & 11 & $0^+$ & 39.13 \\
    \hline
  \end{tabular}
  \caption{The sequential number, the $K$-value, and the squared 
  overlap value are shown for the SDs which dominate in the wave 
  function of the $0_3^+$m $2_4^+$ and $4_2^+$..}
  \label{over_chain}
\end{table}

\begin{figure}[Htb]
  \begin{center}
     \includegraphics[]{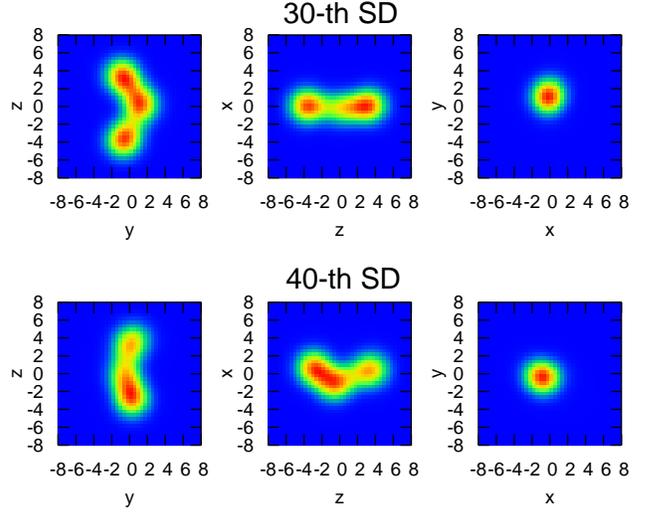}
  \end{center}
  \caption{Contour plots of the density distributions of the 30th and the 
               40th SDs, which are the major components of the $0_3^+$, $2_4^+$, 
               and $4_2^+$ states.}
  \label{dens_linear}
\end{figure}

As seen in Sec. \ref{s:trans}, $0^+_3$, $2_4^+$ and $4_2^+$ states are 
connected by the intense $B(E2)$ values. In Table~\ref{over_chain}, the sequential 
numbers of the SD which have large overlap values with the wave function of
these states are shown. The 30th, the 40th, and the 
19th SDs are commonly included in the three states. We show in 
Fig.~\ref{dens_linear} the density distributions of the 30th and 
the 40th SDs. They clearly show a bended linear-chain structure of 
three $\alpha$ particles.

\section{Mixing of Three-alpha Configurations}
\label{s:mix_conf}

\begin{figure*}[Htb]
  \begin{center}
     \includegraphics[]{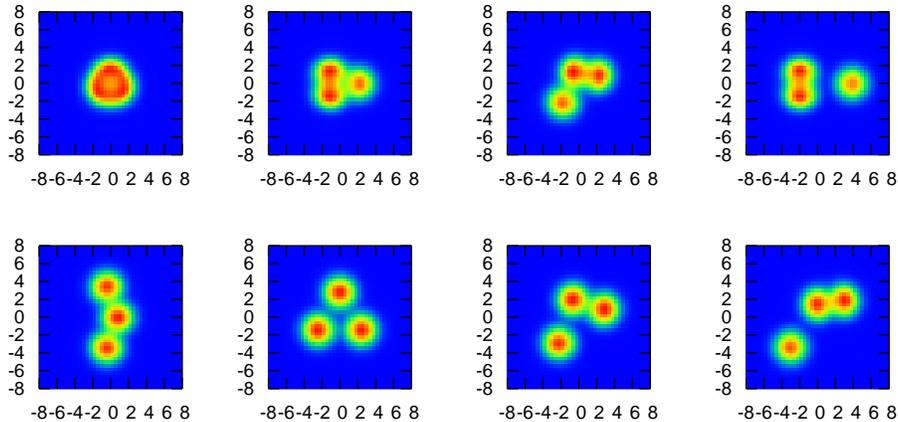}
  \end{center}
  \caption{Density distributions of some SDs out of 31 SDs 
    which are used in Ref.~\cite{EUega1979}.
    Unit of vertical and horizontal axes are fm. }
  \label{wf_cluster} 
\end{figure*}

Some of the present results in Sec.~\ref{s:results} are found to be
qualitatively different from those of AMD and microscopic 
cluster models. For example, the radius of the $0_2^+$ state
is much smaller in our calculation. The charge form factor at large
momentum transfer is described much better by other theories than ours.
These facts may indicate that the imaginary-time propagation
may not sufficiently produce 
a certain class of $\alpha$-cluster wave functions.
In order to check whether explicit inclusion of $\alpha$-cluster 
configurations bring large changes in the current results,
we perform configuration-mixing calculations
including the wave functions similar to those employed 
in the microscopic cluster model of Ref.~\cite{EUega1979}.

The 31 SDs of the $\alpha$-cluster wave functions are
used in the GCM calculation in Ref.~\cite{EUega1979}.
We place the $\alpha$-particle wave functions
at the same positions as those of Ref.~\cite{EUega1979}. 
In Ref.~\cite{EUega1979}, the single-particle wave functions of 
the SDs are the Gaussian wave packets. 
Instead of the Gaussian wave packet, we employ the HF
orbitals of the $\alpha$ particles.
In Fig.~\ref{wf_cluster}, we show density distributions of 
selected SDs among those 31 SDs.
 
\begin{figure}[Htb]
\includegraphics[width=85mm]{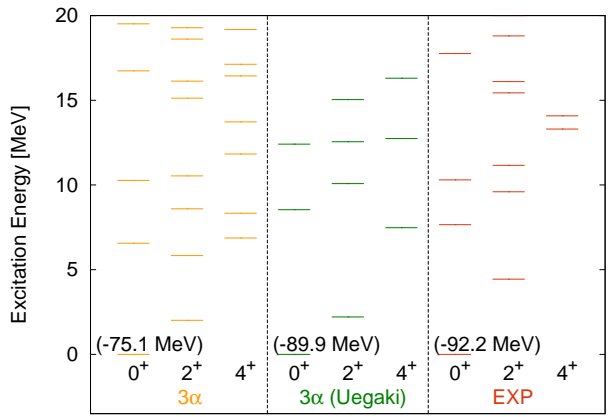}
\caption{Energy levels of $^{12}$C employing  3$\alpha$ SDs with 
Skyrme SLy4 interaction (left), the results of GCM
  calculation of Ref.~\cite{EUega1979} (center) and experiments (right). }
\label{f:cluster_energy}
\end{figure}

In Fig.~\ref{f:cluster_energy}, we show excitation spectra from 
configuration mixing calculations using the 31 SDs. The left panel 
shows our calculation using SLy4 interaction.
The middle panel shows the GCM calculation using Volkov No. I force
\cite{EUega1979}.
The results for the ground rotational band are similar to each other.
In fact, in both calculations, the moment of inertia is too large.
The $0_2^+$ state appears at around 
7 MeV in our calculation, slightly lower than that of Ref.~\cite{EUega1979}.

In the parentheses in Fig.~\ref{f:cluster_energy}, we show the 
calculated binding energies in the ground state. The absolute values of the
binding energies is very different between our calculation and
that of Ref.~\cite{EUega1979}. In our calculation using SLy4 interaction,
the binding energy is -75.1 MeV and is much smaller than the value
shown in Fig.~\ref{f:ene_dif}.
A major part of the difference comes from the
spin-orbit interaction which contributes
little in the calculation using the alpha-cluster wave functions only.

\begin{figure}[Htb]
  \includegraphics[width=90mm]{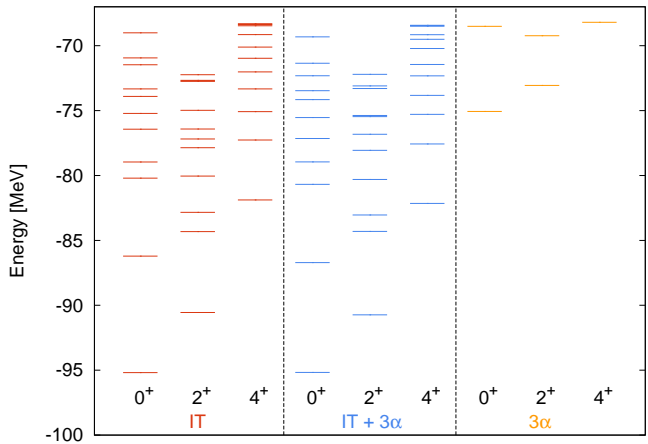}
  \caption{Energy levels of $^{12}$C in the configuration mixing 
  calculation with the SDs constructed by the imaginary-time 
  evolution (IT), and 3$\alpha$, and all of
  these configurations (IT+3$\alpha$). See text for details.}
  \label{f:energy_all} 
\end{figure}

We next perform a configuration mixing calculation employing
both the 50 SDs prepared by the imaginary-time method and the
31 SDs of three-$\alpha$ configuration.
In Fig.~\ref{f:energy_all}, we compare the three calculations:
the configuration
mixing calculation using 50 SDs prepared by the imaginary time
method (IT), the configuration mixing calculation using 31 SDs
of three-$\alpha$ configuration (3$\alpha$), and the configuration 
mixing calculation using both (IT+3$\alpha$). The calculation
labeled by three-$\alpha$ is the same as that shown in the left part
of Fig. \ref{f:energy_all}, except that the total energies 
are plotted here.

After mixing both configurations of the imaginary-time and the
three-$\alpha$, we find the results are very close to the calculation
using the imaginary-time configurations only.
Namely, 31 SDs of the three-$\alpha$ wave functions
do not mix with those prepared by the imaginary-time method.
This is due to the large energy difference between those two
sets of configurations.

In the calculation using configurations generated by the imaginary-time
method, the contribution of the spin-orbit interaction
to the binding energy is as large as 17 MeV with SLy4.
This large energy gain is missing in the pure three-$\alpha$ 
configurations.

In Table \ref{f:radii_summary}, we show the calculated radii and 
the $M(E0)$ transition strength. Using the 31 SDs of three-$\alpha$ 
wave functions, our calculation gives large values for both the 
$0_1^+$ and $0_2^+$ states. The radius of the $0_2^+$ state is 
3.31 fm, close to the value by the GCM calculation, 3.4 fm. 
The $M(E0)$ value is also large, 8.72$e{\rm fm}^2$, even larger than 
the three-$\alpha$ GCM calculation \cite{EUega1979}. 
However, in the configuration mixing calculation using both configurations, 
our calculated values are very close to the calculation using
the 50 SDs prepared by the imaginary-time method. This result is
consistent with the fact that the energy spectra is very little affected
by adding the three-$\alpha$ wave functions.

\begin{table*}[Htb]
  \begin{tabular}{c|c|cccccc|c}
    \hline\hline
    & EXP & IT  & IT + 3$\alpha$ & 3$\alpha$ & 3$\alpha$(Uegaki)\\
    \hline
    radius$(0_1^+)$                & 2.31$\pm$0.02 & 2.53 & 2.54 & 2.80 & 2.40\\
    radius$(0_2^+)$                &               & 2.76 & 2.73 & 3.31 & 3.40\\
    $M(E0;0_2^+\rightarrow0_1^+)$  & 5.4$\pm$ 0.2  & 4.57 & 4.13 & 8.72 & 6.6 \\
    \hline 
  \end{tabular}
  \caption{Radii (fm) and $M(E0)$ ($e{\rm fm}^2$) calculated in various 
  model spaces. Results of GCM calculation \cite{EUega1979} is also 
  shown. See text for details.}
  \label{f:radii_summary}
\end{table*}  

\section{SUMMARY}
\label{s:summary}
We have investigated structure of the $^{12}$C nucleus employing
a configuration-mixing approach with Skyrme interaction. 
In this approach, we first generate a number of Slater determinants 
using the imaginary-time method \cite{SShinohara2006}
starting from initial Slater determinants prepared in a stochastic way.
These Slater determinants show various shapes and clustering.
They are projected on parity and angular momentum,
then, are superposed by the configuration-mixing calculation.
We have generated several sets of Slater determinants and
compare the results with the different sets, to quantify the
reliability of the calculation.
A few low-lying states for each parity and angular momentum 
are well converged with small variance among the different 
sets of the Slater determinants. This fact indicates that the present 
calculation provides unique and convergent results for the ground
and a few low-lying excited states, once the effective 
nucleon-nucleon force, the Skyrme interaction in the present calculation,
is given.

Our calculation reasonably reproduces the overall features of
the structure of $^{12}$C.
The energies and the $B(E2)$ transition strength of the ground state
rotational band are well described.
The lowest excited states of negative parity, $1^-$, $2^-$, and $3^-$, 
are also reasonably described, although the excitation energies are 
slightly too high. The Slater determinants which dominate in these states
show three-$\alpha$ structure.
Our calculation also reproduces the excitation energy of the Hoyle ($0^+_2$)
state reasonably.
This state is found to be described by superposition of many Slater 
determinants, consistent with the former cluster-model calculations.
However, the radius of the $0_2^+$ state is calculated to be
significantly smaller than those.
The energy gain associated with the spin-orbit interaction in the
present method seems to be responsible for this difference.
The three-$\alpha$ liner-chain structure appears at around 15 MeV 
excitation, forming a rotational band.

The success for the description of $^{12}$C nucleus reported in this
paper clearly shows that the present approach is promising for
systematic description of various many-body correlations including
clustering in light nuclei. 

\begin{acknowledgments}

This work is supported by JSPS Kakenhi Grant No. 20105003 and 21340073.
It is also supported by SPIRE Field 5, MEXT, Japan.
Numerical calculations for the present work have been 
carried out on T2K-Tsukuba System at Center for 
Computational Sciences in University of Tsukuba and 
SR16000 at YITP in Kyoto University.

\end{acknowledgments}

\bibliography{draft15}

\end{document}